\documentclass[conference]{IEEEtran}
\ifCLASSOPTIONcompsoc
  \usepackage[nocompress]{cite}
\else
  \usepackage{cite}
\fi
\ifCLASSINFOpdf
  \usepackage[pdftex]{graphicx}
  \DeclareGraphicsExtensions{.pdf,.jpeg,.png}
\else
\fi

\usepackage{url}

\usepackage{enumitem}

\begin{document}
\title{Architectures for High Performance Computing and Data Systems using \\ Byte-Addressable Persistent Memory}

\author{\IEEEauthorblockN{Adrian Jackson\IEEEauthorrefmark{1}, Mark Parsons\IEEEauthorrefmark{2}, Mich\`{e}le Weiland\IEEEauthorrefmark{3}}
\IEEEauthorblockA{\textit{EPCC, The University of Edinburgh} \\
Edinburgh, United Kingdom \\
\IEEEauthorrefmark{1}a.jackson@epcc.ed.ac.uk, \IEEEauthorrefmark{2}m.parsons@epcc.ed.ac.uk, \IEEEauthorrefmark{3}m.weiland@epcc.ed.ac.uk}
\IEEEauthorblockN{Bernhard Hom{\"o}lle\IEEEauthorrefmark{4}}
\IEEEauthorblockA{\textit{SVA System Vertrieb Alexander GmbH} \\
Paderborn, Germany  \\
\IEEEauthorrefmark{4} Bernhard.homoelle@sva.de}}

\maketitle

\begin{abstract}
  Non-volatile, byte addressable, memory technology with performance close to main memory promises to revolutionise computing systems
  in the near future. Such memory technology provides the potential for extremely large memory regions (i.e. $>$ 3TB per server), very high
  performance I/O, and new ways of storing and sharing data for applications and workflows. This paper outlines an architecture that has
  been designed to exploit such memory for High Performance Computing and High Performance Data Analytics systems, along with descriptions
  of how applications could benefit from such hardware.
\end{abstract}


\section{Introduction}\label{sec:introduction}
\IEEEPARstart{T}{here} are a number of new memory technologies that are impacting, or
likely to impact, computing architectures in the near future. One example of such a technology
is so called high bandwidth memory, already featured today on Intel's latest many-core processor,
the Xeon Phi Knights Landing \cite{Sodani:hotchips}, and NVIDIA's latest GPU, Volta \cite{nvidia:volta}. These contain MCDRAM \cite{Sodani:hotchips}
and HBM2 \cite{jun:hbm} respectively, memory technologies built with traditional DRAM hardware but
connected with a very wide memory bus (or series of buses) directly to the processor to provide very high memory bandwidth when compared
to traditional main memory (DDR channels).

This has been enabled, in part, by the hardware trend for incorporating memory controllers and memory controller hubs directly
onto processors, enabling memory to be attached to the processor itself rather than through the motherboard and
associated chipset. However, the underlying memory hardware is the same, or at least very similar, to the traditional volatile DRAM memory
that is still used as main memory for computer architectures, and that remains attached to the motherboard rather than the processor.

Non-volatile memory, i.e. memory that retains data even after power is turned off,
has been exploited by consumer electronics and computer systems for many years. The flash memory
cards used in cameras and mobile phones are an example of such hardware, used for data storage. More
recently, flash memory has been used for high performance I/O in the form of Solid State Disk (SSD) drives,
providing higher bandwidth and lower latency than traditional Hard Disk Drives (HDD).

Whilst flash memory can provide fast input/output (I/O) performance for computer systems, there are some draw backs. It has limited
endurance when compare to HDD technology, restricted by the number of modifications a memory cell can undertake and thus
the effective lifetime of the flash storage\cite{ssd:wear}. It is often also more expensive than other storage technologies. However,
SSD storage, and enterprise level SSD drives, are heavily used for I/O intensive functionality in large scale computer
systems because of their random read and write performance capabilities.

Byte-addressable random access persistent memory (B-APM), also known as storage class memory (SCM), NVRAM or NVDIMMs, exploits a new generation of non-volatile memory hardware
that is directly accessible via CPU load/store operations, has much higher durability than standard flash memory, and much higher
read and write performance.

High-performance computing (HPC) and high-performance data analytics (HPDA) systems currently have different hardware and
configuration requirements. HPC systems generally require very high floating-point performance, high memory bandwidth, and
high-performance networks, with a high-performance filesystem for production I/O and possibly a larger filesystem for long term data storage.
However, HPC applications do not generally have large memory requirements (although there are some exceptions to this) \cite{turner:memory}. HPDA systems
on the other hand often do have high memory capacity requirements, and also require a high-performance filesystem to
enable very large amounts of data to be read, processed, and written.

B-APM, with its very high performance I/O characteristics, and vastly increased capacity (compared to volatile memory), offers a
potential hardware solution to enable the construction of a compute platform that can support both types of use case, with
high performance processors, very large amounts of B-APM in compute nodes, and a high-performance network, providing a
scalable compute, memory, and I/O system.

In this paper, we outline the systemware and hardware required to provide such a system. We start by describing persistent memory, and the functionality it
provides, in more detail in section \ref{sec:pm}. In section \ref{sec:opportunities} we discuss how B-APM could be exploited for
scientific computation or data analytics. Following this we outline our proposed hardware and systemware
architectures in sections \ref{sec:hardware} and \ref{sec:systemware}, describe how applications could benefit from such a system in section \ref{sec:using}.
We finish by discussing related work in section \ref{sec:related} summarising the paper in the final section.


\section{Persistent Memory}\label{sec:pm}
B-APM takes new non-volatile memory technology and packages it in the same form factor (i.e. using the same connector and dimensions)
as main memory (SDRAM DIMM form factor). This allows B-APM to be installed and used alongside DRAM based main memory, accessed through the same memory controller. 

As B-APM is installed in a processor’s memory channels, applications running on the system can access B-APM directly as if it was main memory,
including true random data access at byte or cache line granularity. Such an access mechanism is very different to the traditional block
based approaches used for current HDD or SSD devices, which generally requires I/O to be done using blocks of data (i.e. 4KB of data
written or read in one operation), and relies on expensive kernel interrupts.

The B-APM technology that will be the first to market is Intel and Micron’s 3D XPoint\texttrademark memory \cite{hady:3dxpoint}. The performance of this, byte-addressable,
B-APM, is projected to be lower than main memory (with a latency $\sim$5-10x that of DDR4 memory when connected to the same memory channels),
but much faster than SSDs or HDDs. It is also projected to be of much larger capacity than DRAM, around 2-5x denser (i.e. 2-5x
more capacity in the same form factor).

\subsection{Data Access}
This new class of memory offers very large memory capacity for servers, as well as long term very high performance persistent storage within the memory
space of the servers, and the ability to undertake I/O (reading and writing storage data) in a new way. Direct access (DAX) from applications to
individual bytes of data in the B-APM is very different from the block-oriented way I/O is currently implemented. B-APM has the potential to enable synchronous,
byte level I/O, moving away from the asynchronous block-based file I/O applications currently rely on. In current asynchronous I/O user applications
pass data to the operating system (OS) which then use driver software to issue an I/O command, putting the I/O request into a queue on a hardware controller.
The hardware controller will process that command when ready, notifying the OS that the I/O operation has finished through an interrupt to the device driver.

B-APM can be accessed simply by using a load or store instruction, as with any other memory operation from an application or program.
If the application requires persistence, it must flush the data from the volatile CPU caches and ensure that the same data has also arrived on the non-volatile medium.
There are cache flush commands and fence instructions. To keep the performance for persistence writes, new cache flush operation have been introduced. Additionally, write buffers
in the memory controller may be protected by hardware through modified power supplies (such as those supporting asynchronous DRAM refresh \cite{nvdimm:snia}).

With B-APM providing much lower latencies than external storage devices, the traditional I/O block access model, using interrupts, becomes
inefficient because of the overhead of context switches between user and kernel mode (which can take thousands of CPU cycles\cite{contextswitch}). Furthermore, with
B-APM it becomes possible to implement remote persistent access to data stored in the memory using RDMA technology over a suitable
interconnect. Using high performance networks can enable access to data stored in B-APM in remote nodes faster than accessing local high
performance SSDs via traditional I/O interfaces and stacks inside a node.

Therefore, it is possible to use B-APM to greatly improve I/O performance within a server, increase the memory capacity of a server,
or provide a remote data store with high performance access for a group of servers to share. Such storage hardware can also be scaled
up by adding more B-APM memory in a server, or adding more nodes to the remote data store, allowing the I/O performance of a system to scale as required.
However, if B-APM is provisioned in the servers, there must be software support for managing data within the B-APM.
This includes moving data as required for the jobs running on the system, and providing the functionality to let applications
run on any server and still utilise the B-APM for fast I/O and storage (i.e. applications should be able to access B-APM in remote nodes
if the system is configured with B-APM only in a subset of all nodes).

As B-APM is persistent, it also has the potential to be used for resiliency, providing backup for data from active applications, or
providing long term storage for databases or data stores required by a range of applications. With support from the systemware,
servers can be enabled to handle power loss without experiencing data loss, efficiently and transparently recovering from power
failure and resuming applications from their latest running state, and maintaining data with little overhead in terms of performance.

\subsection{B-APM modes of operation}

Ongoing developments in memory hierarchies, such as the high bandwidth memory in Xeon Phi manycore processors or NVIDIA GPUS,
have provided new memory models for programmers and system designers/implementers. A common model that has been proposed includes
the ability to configure main memory and B-APM in two different modes: Single-level and Dual-level memory \cite{multilevel:intel}.

\begin{figure}[!t]
\centering
\includegraphics[width=2.5in]{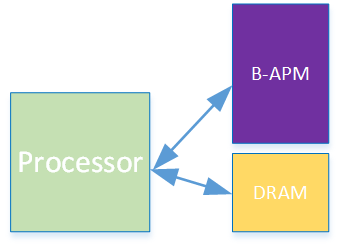}
\caption{Single-level memory (SLM) configuration using main memory and B-APM}
\label{1lm_pic}
\end{figure}

Single-level memory, or SLM, has main memory (DRAM) and B-APM as two separate memory spaces, both accessible by applications, as outlined
in Figure \ref{1lm_pic}. This is very similar to the Flat Mode \cite{mcdram:colfax} configuration of the high bandwidth, on-package, MCDRAM in current Intel
Knights Landing processor. The DRAM is allocated and managed via standard memory API's such as {\it malloc} and represent the OS visible
main memory size. The B-APM will be managed by programming APIs and present the non-volatile part of the system memory. Both will allow
direct CPU load/store operations. In order to take advantage of B-APM in SLM mode, systemware or applications have to be adapted to
use these two distinct address spaces.

\begin{figure}[!t]
\centering
\includegraphics[width=2.5in]{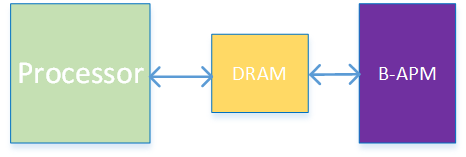}
\caption{Dual-level memory (DLM) configuration using main memory and B-APM}
\label{2lm_pic}
\end{figure}

Dual-level memory, or DLM, configures DRAM as a cache in front of the B-APM, as shown in Figure \ref{2lm_pic}.
Only the memory space of the B-APM is available to applications, data being used is stored in DRAM, and
moved to B-APM when no longer immediately required by the memory controller (as in standard CPU caches).
This is very similar to the Cache Mode \cite{mcdram:colfax} configuration of MCDRAM on KNL processors.

This mode of operation does not require applications to be altered to exploit the
capacity of B-APM, and aims to give memory access performance at main memory speeds whilst providing
access to the large memory space of B-APM. However, exactly how well the main memory cache performs will
depend on the specific memory requirements and access pattern of a given application.
Furthermore, persistence of the B-APM contents cannot be longer guaranteed, due to the volatile DRAM
cache in front of the B-APM, so the non-volatile characteristics of B-APM are not exploited.

\subsection{Non-volatile memory software ecosystem}

The Storage Networking Industry Association (SNIA) have produced a software architecture for B-APM
with persistent load/store access, formalised in the Linux Persistent Memory Development Kit (PMDK) \cite{pmem} library.

\begin{figure}[!t]
\centering
\includegraphics[width=2.5in]{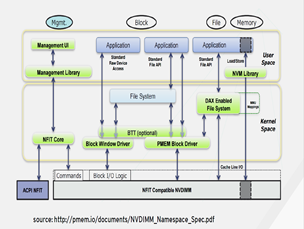}
\caption{PMDK software architecture}
\label{pmem_pic}
\end{figure}

This approach re-uses the naming scheme of files as traditional persistent entities and map the B-APM regions
into the address space of a process (similar to memory mapped files in Linux). Once the mapping has been done,
the file descriptor is no longer needed and can be closed. Figure \ref{pmem_pic} outlines the PMDK software architecture.

\section{Opportunities for exploiting B-APM for computational simulations and data analytics}\label{sec:opportunities}

Reading data from and writing it to persistent storage is usually not the most time consuming part of computational simulation applications.
Analysis of common applications from a range of different scientific areas shows that around 5-20\% of runtime for applications is involved
in I/O operations \cite{layton:IO}\cite{Luu:IO}. It is evident that B-APM can be used to improve I/O performance for applications by replacing
slower SSDs or HDDs in external filesystems. However, such a use of B-APM would be only an incremental improvement in I/O performance, and would
neglect some of the significant features of B-APM that can provide performance benefits for applications.

Firstly, deploying B-APM as an external filesystem would require provisioning a filesystem on top of the B-APM hardware. Standard storage devices
require a filesystem to enable data to be easily written to or read from the hardware. However, B-APM does not require such functionality, and data
can be manipulated directly on B-APM hardware simply through load store instructions. Adding the filesystem and associated interface guarantees
(i.e. POSIX interface \cite{POSIX}), adds performance overheads that will reduce I/O performance on B-APM.

\begin{figure}[!t]
\centering
\includegraphics[width=2.5in]{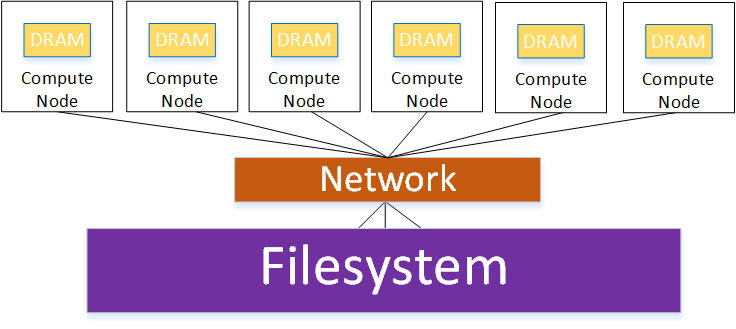}
\caption{Current external storage for HPC and HPDA systems}
\label{external_pic}
\end{figure}

Secondly, an external B-APM based filesystem would require all I/O operations to be performed over a network connection, as current filesystems
are external to compute nodes (see Figure \ref{external_pic}). This would limit the maximum performance of I/O to that of the network between compute nodes and
the nodes the B-APM is hosted in.

\begin{figure}[!t]
\centering
\includegraphics[width=2.5in]{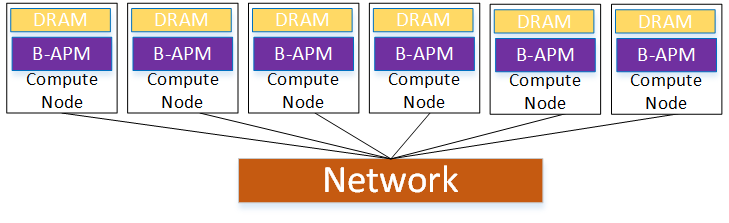}
\caption{Internal storage using B-APM in compute nodes for HPC and HPDA systems}
\label{internal_pic}
\end{figure}

Our vision \cite{weiland:scm} for exploiting B-APM for HPC and HPDA systems is to incorporate the B-APM into the compute nodes, as outlined in Figure \ref{internal_pic}. This architecture allows
applications to exploit the full performance of B-APM within the compute nodes they are using, by giving them the ability to access B-APM through load/store at
byte-level granularity, as opposed to block based, asynchronous I/O in traditional storage devices. Incorporating B-APM into compute nodes also has the benefit that
I/O capacity and bandwidth can scale with the number of compute nodes in the system. Adding more compute nodes will increase the amount of B-APM in the system and add more aggregate bandwidth to I/O/B-APM operations. 

For example, current memory bandwidth of a HPC system scales with the number of nodes used. If we assume an achievable memory bandwidth per node of 100GB/s, then it follows that a system with 10 nodes has the potential to provide 1TB/s of memory bandwidth for a distributed application, and a system with 10000 nodes can provide 1PB/s of memory bandwidth. If an application is memory bandwidth bound and can parallelise across nodes then scaling up nodes like this clearly has the potential to improve performance. For B-APM in nodes, and taking 3D XPoint\texttrademark as an example, if we assume 20GB/s of memory bandwidth per node (5x less than the volatile memory bandwidth), then scaling up to 10 nodes provides 200GB/s of (I/O) memory bandwidth and 10000 nodes provides 200TB/s of (I/O) memory bandwidth. For comparison, the Titan system at ORNL has a Lustre file system with 1.4TB/s of bandwidth \cite{ornl:hardware} and they are aiming for 50TB/s of burst buffer\cite{dayley:burst} I/O by 2024 \cite{ornl:req}.
 
Furthermore, there is the potential to optimise not only the performance of a single application, but rather the performance of a whole scientific workflow,
from data preparation through simulations, to data analysis and visualisation. Optimising full workflows by sharing data between different stages or steps
in the workflow has the potential to completely remove, or greatly reduce, data movement/storage costs for large parts of the workflow altogether. Leaving data
in-situ on B-APM for other parts of the workflow can significantly improve the performance of analysis and visualisation steps at the same time as reducing I/O costs
for the application when writing the data out.

The total runtime of an application can be seen as the sum of its compute time, plus the time spent in I/O.  Greatly reduced I/O costs therefore also have the
beneficial side effect of allowing applications to perform more I/O within the same total cost of the overall simulation. This enables applications to maintain
I/O costs in line with current behaviour whilst being able to process significantly more data. Furthermore, for those applications for which I/O does take up
a large portion of the run time, including data analytics applications, B-APM has the potential to significantly reduce runtime.

\subsection{Potential caveats}

However, utilising internal storage is not without drawbacks. Firstly, the benefit of external storage is that there is a single namespace and location for
compute nodes to use for data storage and retrieval. This means that applications running on the compute nodes can access data trivially as it is stored
externally to the compute nodes. With internal storage, this guarantee is not provided. Data written to B-APM is local to specific compute nodes. It is therefore
necessary for applications to be able to manage and move data between compute nodes, as well as to external data storage, or for some systemware components to
undertake this task.

Secondly, B-APM may be expensive to provision in all compute nodes. It may not be practical to add the same amount of B-APM to all compute nodes, and systems may be
constructed with islands of nodes with B-APM, and islands of nodes without B-APM. Therefore, application or systemware functionality to enable access to remote B-APM
and to exploit/manage asymmetric B-APM configurations will be required. Both these issues highlight the requirement for an integrated hardware and software (systemware)
architecture to enable efficient and easy use of this new memory technology in large scale computational platforms.

\section{Hardware architecture}\label{sec:hardware}

As 3D Xpoint\textsuperscript{TM} memory, and other B-APM when it becomes available, is designed to fit into standard memory form factors and be utilised using the same
memory controllers that main memory exploit, the hardware aspect of incorporating B-APM into a compute server or system is not onerous. Standard HPC and HPDA systems comprise
a number of compute nodes, connected together with a high performance network, along with login nodes and an
external filesystem. Inside a compute node there are generally 2 or more multicore processors, connected to a shared motherboard, with associated
volatile main memory provided for each processor. One or more network connections are also required in each node, generally connected to the
PCIe bus on the motherboard.

To construct a compute cluster that incorporates B-APM all that is required is a processor and associated memory controller that support such memory.
Customised memory controllers are required to intelligently deal with the variation in performance between B-APM and traditional main memory
(i.e. DDR). For instance, as B-APM has a higher access latency than DDR memory it would impact performance if B-APM accesses were blocking, i.e. if the
memory controller could not progress DDR accesses whilst an B-APM access was outstanding. However, other than modifying the memory controller to support such
variable access latencies, it should be possible to support B-APM in a standard hardware platform, provided that sufficient capacity for memory is provided. 

\begin{figure}[!t]
\centering
\includegraphics[width=2.5in]{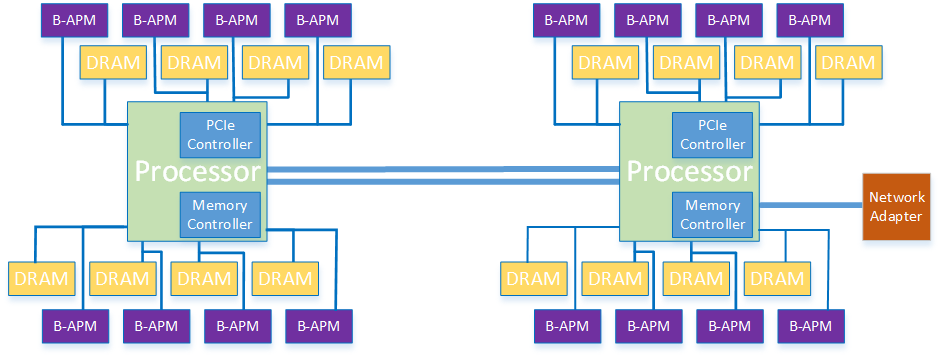}
\caption{Compute node hardware architecture}
\label{node_pic}
\end{figure}

Given both DRAM and B-APM are connected through the same memory controller, and memory controllers have a number of memory channels, it is also important to consider the balance of DRAM and B-APM attached to a processor. If we assume a processor has 6 memory channels, to get full DRAM bandwidth we require at least one DRAM DIMM per memory channel. Likewise, if we want full B-APM bandwidth we need a B-APM DIMM per memory channel. Assuming that a memory channel can support are two DIMM slots, this leads us to a configuration with 6 DRAM DIMMs and 6 B-APM DIMMs per processor, and double that with two processors per node.  This configuration is also desirable to enable the DLM configuration, as DLM requires DRAM available to act as a cache for B-APM, meaning at least a DRAM DIMM is required per memory controller.

Pairing DRAM and B-APM DIMMs on memory channels is not required for all systems, and it should be possible to have some memory channels with no B-APM installed, or some memory channels with no DRAM DIMMs installed. However, if DLM mode is required on a system, it is sensible to expect that at least one DRAM DIMM must be installed per memory controller in addition to B-APM. Future system design may consider providing more than two DIMM slots per memory channel to facilitate systems with different memory configurations (i.e. more B-APM than DRAM DIMMs or memory controllers enabling full B-APM population of memory channels).

The proposed memory configuration allows us to investigate possible system configurations using B-APM memory. Table \ref{tab_scale_nodes} outlines different systems, assuming 3TB of B-APM per node, with a node capable of 2TFlop/s compute performance. These projections are achievable with existing processor technology, and demonstrate that very significant I/O bandwidth can be provided to match the compute performance achieved when scaling to very large numbers of nodes.

\begin{table}[ht]
\caption{B-APM enabled system configurations}
\begin{center}
\begin{tabular}{|l|c|c|c|}
\hline
\textbf{Number of} & \textbf{Compute} & \textbf{B-APM Capacity}& \textbf{B-APM Storage I/O}\\
\textbf{nodes} & \textbf{(PFlop/s)} & \textbf{(PB)}& \textbf{Bandwidth (TB/s)}\\
\hline
1 & 0.002 & 0.003 & 0.02 \\
\hline
768 & 1.5 & 2.3 & 15 \\
\hline
3072 & 6 & 9 & 61 \\
\hline
24576 & 49 & 73 & 491 \\
\hline
196608 & 393 & 589 & 3932 \\
\hline
\end{tabular}
\label{tab_scale_nodes}
\end{center}
\end{table}

Integrating new memory technology in existing memory channels does mean that providing sufficient locations for both main memory and B-APM to be
added is important. Depending on the size of B-APM and main memory technology available, sufficient memory slots must be provided per processor to
allow a reasonable amount of both memory types to be added to a node. Therefore, we are designing our system around a standard compute node architecture
with sufficient memory slot provision to support large amounts of main memory and B-APM as shown in Figure \ref{node_pic}.

Another aspect, which we are not focusing on in the hardware architecture, is data security. As B-APM enables data to be retained inside compute nodes,
ensuring the security of that data, and ensuring that it cannot be accessed by users or applications that are not authorised to access the data is
important. The reason that we are not focusing on this in the hardware architecture is because this requirement can be addressed in software, but
it may also be sensible to integrate encryption directly in the memory hardware, memory controller, or processor managing the B-APM.

\section{Systemware architecture}\label{sec:systemware}

Systemware implements the software functionality necessary to enable users to easily and efficiently utilise the system. We have designed a
systemware architecture that provides a number of different types of functionality, related to different methods for exploiting B-APM for large
scale computational simulation or data analytics.

From the hardware features B-APM provides, our analysis of current HPC and HPDA applications and functionality they utilise, and our
investigation of future functionality that may benefit such applications, we have identified a number of different kinds of functionality
that the systemware architecture should support:
\begin{enumerate}
  \item Enable users to be able to request systemware components to load/store data in B-APM prior to a job starting, or after a job has completed. This
  can be thought of as similar to current burst buffer technology. This will allow users to be able to exploit B-APM without changing their applications.
  \item Enable users to directly exploit B-APM by modifying their applications to implement direct memory access and management. This offers users the
  ability to access the best performance B-APM can provide, but requires application developers to undertake the task of programming for B-APM themselves,
  and ensure they are using it in an efficient manner.
  \item Provide a filesystem built on the B-APM in compute nodes. This allows users to exploit B-APM for I/O operations without having to fundamentally change
  how I/O is implemented in their applications. However, it does not enable the benefit of moving away from file based I/O that B-APM can provide.
  \item Provide an object, or key value, store that exploits the B-APM to enable users to explore different mechanisms for storing and accessing data
  from their applications.
  \item Enable the sharing of data between applications through B-APM. For example, this may be sharing data between different components of the same
  computational workflow, or be the sharing of a common dataset between a group of users. 
  \item Ensure data access is restricted to those authorised to access that data and enable deletion or encryption of data to make sure those access
  restrictions are maintained
  \item Provision of profiling and debugging tools to allow application developers to understand the performance characteristics of B-APM, investigate
  how their applications are utilising it, and identify any bugs that occur during development.
  \item Efficient check-pointing for applications, if requested by users.
  \item Provide different memory modes if they are supported by the B-APM hardware.
  \item Enable or disable systemware components as required for a given user application to reduce the performance impact of the systemware to a
  running application, if that application is not using those systemware components.
\end{enumerate}

\begin{figure*}[!t]
\centering
\includegraphics[width=\textwidth]{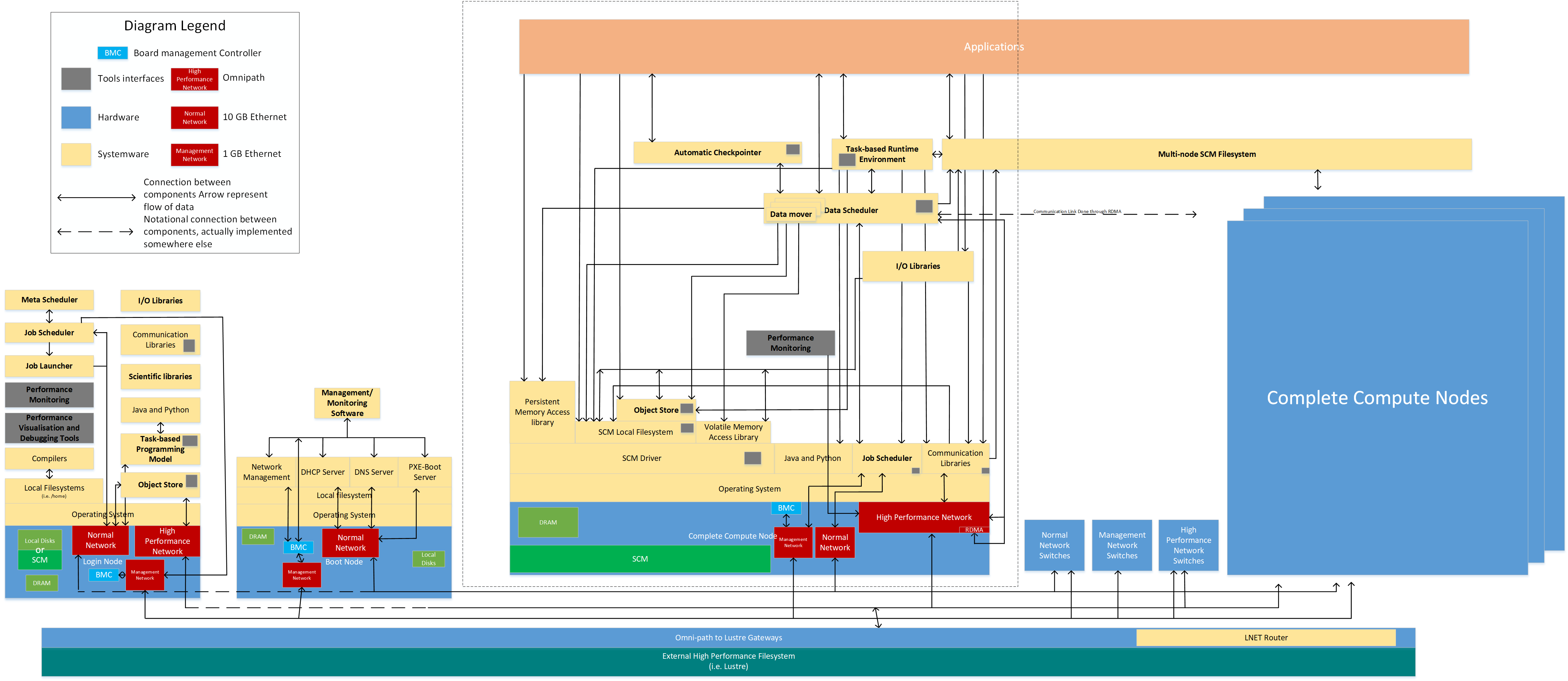}
\caption{Systemware architecture to exploit B-APM hardware in Compute nodes}
\label{systemware_pic}
\end{figure*}

The systemware architecture we have defined is outlined in Figure \ref{systemware_pic}. Whilst this architecture may appear to have a large number of components and significant
complexity, the number of systemware components that are specific to a system that contains B-APM is relatively small. The new or modified components we have identified are
required to support B-APM in a large scale, multi-user, multi-application, compute platforms are as follows:
\begin{itemize}
  \item Job Scheduler
  \item Data Scheduler
  \item Object Store
  \item Filesystem
  \item Programming Environment
\end{itemize}

We will describe these in more detail in the following subsections.

\subsection{Job scheduler}
As the innovation in our proposed system is the inclusion of B-APM within nodes, one of the key components that must support the new hardware resource
is the job scheduler. Job schedulers, or batch systems, are used to manage, schedule, and run user jobs on the shared resource that are the compute nodes.
Standard job schedulers are configured with the number of nodes in a system, the number of cores per node, and possibly the amount of memory or whether
there are accelerators (like GPUs) in compute nodes in a system. They then use this information, along with a scheduling algorithm and scheduling policies
to allocate user job request to a set of compute nodes. Users submit job requests specifying the compute resources required (i.e. number of node or number
of compute cores a job will require) along with a maximum runtime for the job. This information is used by the job scheduler to accurately, efficiently, and fairly
assign applications to resources.

Adding B-APM to compute nodes provides another layer of hardware resource that needs to be monitored and scheduled against by the job scheduler. As data can persist
in B-APM, and one of our target use cases is the sharing of data between applications using B-APM, the job scheduler needs to be extended to both be aware of
this new hardware resource, and to allow data to be retained in B-APM after an individual job has finished. This functionality is achieved through adding
workflow awareness to the job scheduler, providing functionality to allow data to be retained and shared
through jobs participating in the workflow, although not indefinitely\cite{farasarakis:PDSW}. The job scheduler also needs to be able to clean up the B-APM after a job has
finished, ensuring no data is left behind or B-APM resources consumed, unless specifically as part of a workflow.

Furthermore, as the memory system is likely to have different modes of operation, the job scheduler will need to be able to query the current
configuration of the memory hardware, and be able to change configuration modes if required by the next job that will be using a particular set of compute nodes.
We are also investigating new scheduling algorithms, specifically data aware and energy aware scheduling algorithms, to optimise system
efficiency or throughput using B-APM functionality. These will utilise the job scheduler’s awareness of B-APM functionality and compute job data requirements.

\subsection{Data scheduler}
The data scheduler is an entirely new component, designed to run on each compute node and provide data movement and shepherding functionality. 
Much of the new functionality we are implementing to exploit B-APM for users involves moving data to and from B-APM asynchronously (i.e. pre-loading data
before a job starts, or moving data from B-APM after a job finishes). Furthermore, we also require the ability to move data between different nodes
(i.e. in the case that a job runs on a node without B-APM and requires B-APM functionality, or a job runs and needs to access data left on B-APM in a
different node by another job).

To provide such support without requiring users to modify their applications we implement such functionality in the data scheduler component.
This component has interfaces for applications to interact with, and is also interfaced with the job scheduler component on each compute node. Through
these interfaces the data scheduler can be instructed to move data as required by a given application or workflow.

\subsection{Object store}
We recognise the performance and functionality benefits that exploiting new storage technologies can bring to applications. We are therefore investigating the use of object stores, such as DAOS\cite{lofstead:DAOS} and dataClay\cite{marti:dataclay} and are porting them to the hardware architecture we are proposing, i.e. systems with distributed B-APM as the main storage hardware.

\subsection{Filesystems}
As previously discussed, there is a very large pre-existing code base currently exploiting HPC and HPDA systems. The majority of these will
undertake I/O using files through an external filesystem. Therefore, the easiest mechanism for supporting such applications, in the first instance,
is to provide filesystems hosted on the B-APM hardware.

Our architecture provides the functionality for two different types of filesystems using B-APM:
\begin{itemize}
  \item Local, on-node
  \item Distributed, cross-node
\end{itemize}

The local filesystem will provide applications with a space for reading or writing data from/to a filesystem that is separate for each compute node,
i.e. a scratch or \texttt{/tmp} filesystem on each node. This will enable very high performance file I/O, but require applications (or the data scheduler) to manage these files. It will also provide a storage space for files to be loaded prior to a job starting (i.e. similar to burst buffer functionality), or for files that
should be move to an external filesystem when a job has finished. The distributed filesystem will provide functionality similar to current parallel filesystems
(e.g Lustre), except it will be hosted directly on the B-APM hardware and not require external I/O servers or nodes.

\subsection{Programming environment}
Finally, the programming environment, i.e. libraries, compilers, programming languages, communication libraries, used by application needs to be
modified to support B-APM. An obvious example of such requirements is ensuring that common I/O libraries support using B-APM for storage. For instance,
many computational simulation applications use MPI-I/O, HDF5, or NetCDF to undertake I/O operations. Ensuring these libraries can utilise B-APM in some
way to undertake high performance I/O will ensure a wide range of existing applications can exploit the system effectively.

Further modifications, or new functionality, may also be beneficial. For instance, we will deploy a task based programming environment, PyCOMPs\cite{tejedor:pycomps}, which can interact directly with object storage. This will enable us to evaluate whether new parallel programming approaches will enable exploitation of B-APM and
the functionality it provides more easily than adapting existing applications to functionality such as object stores or byte-based data accesses to B-APM.

\section{Using B-APM}\label{sec:using}
To allow a fuller understanding of how a system developed from the architectures we have designed could be used, we discuss some of the possible usage
scenarios in the following text. We outline the systemware and hardware components used by a given use case, and the lifecycle of the data in those components.

\subsection{Filesystems on B-APM}
For this use case we assume an application undertaking standard file based I/O operations, using either parallel or serial I/O functionality. We assume
the distributed, cross-node, B-APM filesystem is used for I/O whilst the application is running, and the external high performance filesystem is used for data storage
after an application has finished. 

\begin{figure}[!t]
\centering
\includegraphics[width=2.5in]{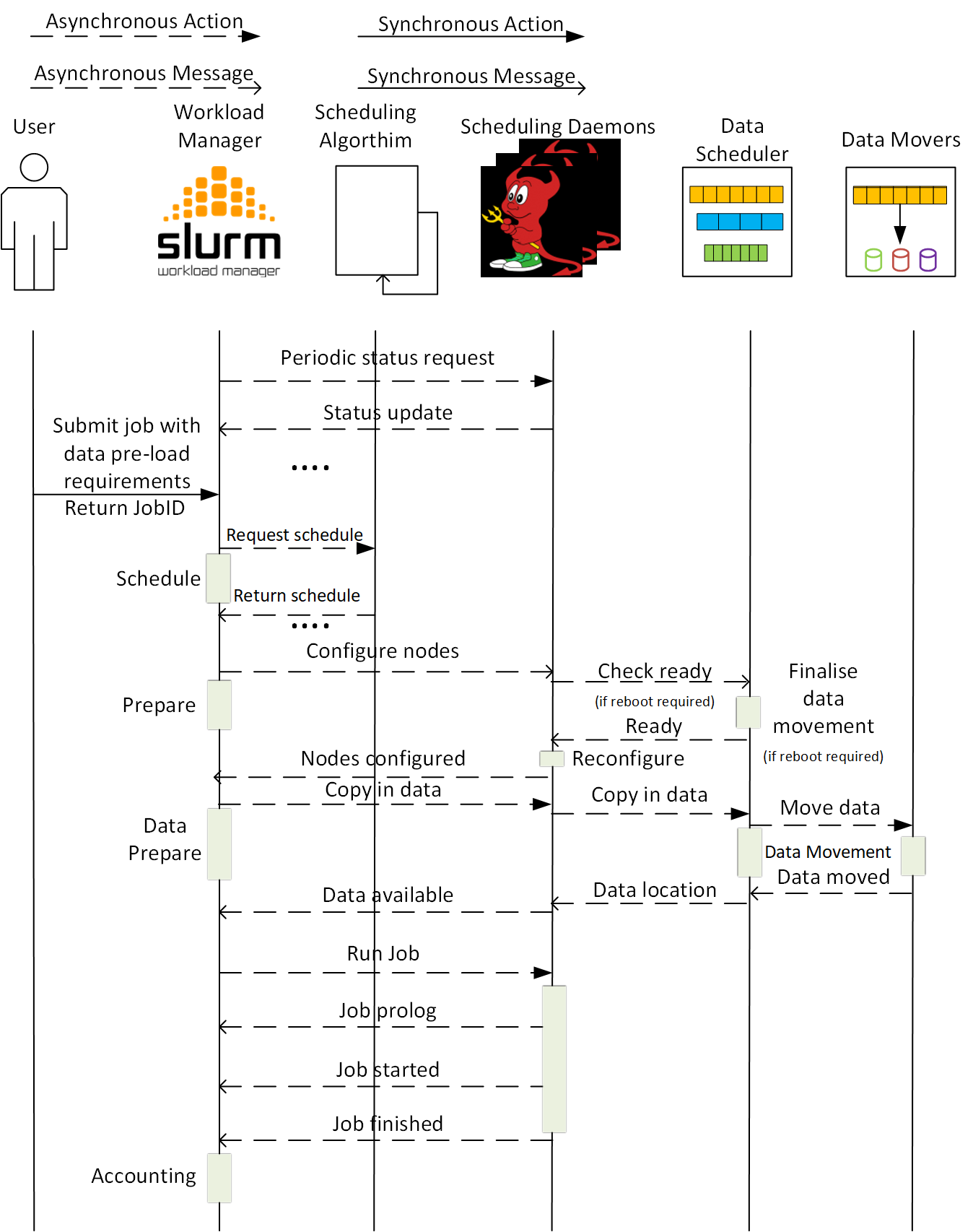}
\caption{Sequence diagram for systemware component use by application requesting data be loaded into distributed B-APM-based filesystem prior to starting}
\label{filesystem_seq_pic}
\end{figure}

The job scheduling request includes a request for files to be pre-loaded on to the distributed B-APM filesystem prior to the job starting. In this use case
we expect the following systemware interaction to occur (also outlined in Figure \ref{filesystem_seq_pic}):
\begin{enumerate}
  \item User job scheduling requests submitted.
    \begin{enumerate}[label=\alph*.]
      \item At this point the application to be run is either stored on the external high performance filesystem or on the local filesystem on the login nodes,
    and the data for the application is stored on the external high performance filesystem.
    \end{enumerate}
  \item Job scheduler allocates resources for an application.
    \begin{enumerate}[label=\alph*.]
      \item The job scheduler ensures that nodes are in SLM mode and that the multi-node B-APM filesystem component is operational on the nodes being
      used by this application.
    \end{enumerate}
  \item Once the nodes that the job has been allocated to are available the data scheduler (triggered by the job scheduler) copies input data from the external
    high performance filesystem to the multi-node B-APM filesystem.
      \begin{enumerate}[label=\alph*.]
        \item This step is optional; an application could read input data directly from the external high performance filesystem, but using the multi-node
        B-APM filesystem will deliver better performance.
      \end{enumerate}
    \item Job Launcher starts the user application on the allocated compute nodes.
    \item Application reads data from the multi-node B-APM filesystem.
    \item Application writes data to the multi-node B-APM filesystem.
    \item Application finishes.
    \item Data Scheduler is triggered and moves data from multi-node B-APM filesystem to the external high performance filesystem.
\end{enumerate}

\section{Related work}\label{sec:related}

There are existing technological solutions that are offering similar functionality to B-APM and that can also be exploited for high performance I/O.
One example is NVMe devices: SSDs that are attached to the PCIe bus and support the NVM Express interface.  Indeed, Intel already has a line of an NVMe device
on the market that use 3D XPoint\texttrademark memory technology, called Intel Optane. Other vendors have a large range of NVMe devices on the market, most of them based on
different variations of Flash technology.

NVMe devices have the potential to provide byte-level storage access, using the {\it PMDK} libraries. A file can be opened and presented as a memory
space for an application, and then can be used directly as memory by that application, removing the overhead of file access (i.e. data access through
file reads and writes) when performing I/O and enabling the development of applications that exploit B-APM functionality.  However, given that NVMe devices are connected
via the PCIe bus, and have a disk controller on the device through which access is managed, NVMe devices do not provide the same level of performance that B-APM offers.
Indeed, as these devices still use block-based data access, fine grained memory operations can require whole blocks of data to be loaded or stored to the device,
rather than individual bytes.

There are a wide range of parallel and high performance filesystems designed to enable high performance I/O from large scale compute clusters\cite{schwan:Lustre}\cite{schmuck:GPFS}\cite{beegfs}.
However, these provide POSIX compliant block based I/O interfaces, which do not offer byte level data access, requiring conversion of data from program
data structures to a flat file format.  Furthermore, whilst it is advantageous that such filesystems are external resources, and therefore can be accessed from any compute node in a cluster,
this means that filesystem performance does not necessarily scale with compute nodes. Such filesystems are specified and provisioned separately from
the compute resource in a HPC or HPDA system. Work has been done to optimise I/O performance of such high performance
filesystems\cite{jian:lustre}\cite{choi:storage}\cite{lin:lustre}\cite{carns:storage}, but they do not address B-APM or new mechanisms
for storing or accessing data without the overhead of a POSIX-compliant (or weakly-compliant) filesystem. 

Another technology that is being widely investigated for improving performance and changing I/O functionality for applications is some form of
object, key value, store\cite{kim:keystore}\cite{lofstead:DAOS}\cite{marti:dataclay}. These provide alternatives to file-based data storage, enabling data to be stored in similar formats or
structures as those used in the application itself. Object stores can start to approach byte level access granularity, however, they require
applications to be significantly re-engineered to exploit such functionality.

We are proposing hardware and systemware architectures in this work that will integrate B-APM into large scale compute clusters, providing
significant I/O performance benefits and introducing new I/O and data storage/manipulation features to applications. Our key goal is to
create systems that can both exploit the performance of the hardware and support applications whilst they port to these new I/O or data
storage paradigms.

Indeed, we recognise that there is a very large body of existing applications and data analysis workflows that cannot immediately be
ported to new storage hardware (for time and resource constraint reasons). Therefore, our aims in this work are to provide a system that
enables applications to obtain best performance if porting work is undertaken to exploit B-APM hardware features, but still allow applications
to exploit B-APM and significantly improve performance without major software changes.

\section{Summary}\label{sec:summary}
This paper outlines a hardware and systemware architecture designed to enable the exploitation of B-APM hardware directly by applications,
or indirectly by applications using systemware functionality that can exploit B-APM for applications. This dual nature of the system provides
support for existing application to exploit this emerging memory new hardware whilst enabling developers to modify applications to best exploit the hardware over time.

The system outlined provides a range of different functionality. Not all functionality will be utilised by all applications, but providing
a wide range of functionality, from filesystems to object stores to data schedulers will enable the widest possible use of such systems.
We are aiming for hardware and systemware that enables HPC and HPDA applications to co-exist on the same platform.

Whilst the hardware is novel and interesting in its own right, we predict that the biggest benefit in such technology will be realised through
changes in application structure and data storage approaches facilitated by the byte-addressable persistent memory that will become routinely
available in computing systems.

In time it could possible to completely remove the external filesystem from HPC and HPDA systems, removing hardware complexity and the energy/cost
associated with such functionality. There is also the potential for volatile memory to disappear from the memory stack everywhere except on the processor
itself, removing further energy costs from compute nodes. However, further work is required to evaluate the impact of the costs of the active systemware
environment we have outlined in this paper, and the memory usage patterns of applications.

Moving data asynchronous to support applications can potentially bring big performance benefits but the impact such functionality has on applications
running on those compute node needs to be investigated. This is especially important as with distributed filesystems or object stores hosted on node distributed B-APM such
in-node asynchronous data movements will be ubiquitous, even with intelligent scheduling algorithms.

\appendices
\section*{Acknowledgements}
The NEXTGenIO project\footnote{www.nextgenio.eu} and the work presented in this paper were funded by the European Union’s Horizon 2020 Research and Innovation programme under Grant Agreement no. 671951. All the NEXTGenIO Consortium members (EPCC, Allinea, Arm, ECMWF, Barcelona Supercomputing Centre, Fujitsu Technology Solutions, Intel Deutschland, Arctur and Technische Universit\"at Dresden) contributed to the design of the architectures.

\ifCLASSOPTIONcaptionsoff
  \newpage
\fi

\IEEEtriggeratref{24}


%

\end{document}